\documentclass[pra,twocolumn,showpacs,superscriptaddress,preprintnumbers,amsmath,amssymb,floatfix]{revtex4}

\usepackage{graphicx}
\usepackage{dcolumn}
\usepackage{bm}
\newcommand\ket[1]{\left\vert#1\right\rangle}
\newcommand\bra[1]{\left\langle#1\right\vert}
\newcommand\mat[1]{\mathbf{#1}}
\newcommand\abs[1]{\left\vert#1\right\vert}

\begin{document}


\title{Trap-induced resonances in controlled collisions of cesium atoms}

\author{Ren\'e Stock}
\email{restock@qis.ucalgary.ca}
\affiliation{{Institute for Quantum Information Science, University of Calgary, Alberta T2N 1N4, Canada}}
\affiliation{{Department of Physics and Astronomy, University of New Mexico, Albuquerque, New Mexico 87131, USA}}
\author{Ivan H. Deutsch}
\affiliation{{Department of Physics and Astronomy, University of New Mexico, Albuquerque, New Mexico 87131, USA}}

\date{\today}


\begin{abstract}
We consider the feasibility of observing a trap-induced resonance [Stock {\it et al.}, Phys. Rev. Lett. {\bf 91}, 183201 (2003)] for the case of two $^{133}$Cs atoms, trapped in separated wells of a polarization-gradient optical lattice, and interacting through a multichannel scattering process. Due to the anomalously large scattering length of cesium dimers, a strong coupling can occur between vibrational states of the trap and a weakly bound molecular state that is made resonant by the ac-Stark shift of the lattice. We calculate the energy spectrum of the two-atom system as a function of the distance between two potential wells by connecting the solutions of the Schr\"{o}dinger equation for the short-range molecular potential to that of the long-range trap in a self-consistent manner. The short-range potential is treated through a multichannel pseudopotential, parametrized by the $K$ matrix, calculated numerically for atoms in free space in a close-coupling approximation. This captures both the bound molecular spectrum as well as the energy-dependent scattering for all partial waves. We establish realistic operating conditions under which the trap-induced resonance could be observed and show that this strong and coherent interaction could be used as a basis for high-fidelity two-qubit quantum logic operations.
\end{abstract}


\maketitle

\section{Introduction}

Ultracold atoms in tightly confining traps are of interest in a variety of physics disciplines, with important applications to the study of quantum degenerate gases and the implementation of quantum information processors~\cite{nature:insight}. At the heart of these systems are two-body atomic interactions whose precise controllability via magnetic Feshbach resonances~\cite{Inouye:98} has opened the door to the exploration of new phenomena such as the dynamic collapse of a Bose-Einstein Condensate (BEC)~\cite{Cornell:02}, the production of ultracold molecules~\cite{Regal:03,Herbig:03}, and the observation of the BEC/BCS crossover~\cite{Greiner:03,Jochim:03,Zwierlein:03}.  Strong confinement and reduced dimensionality can have a large effect on the tunability of the atomic coupling constant~\cite{Olshanii:98, Petrov:00, Blume:02,Bolda:03}. For example, quasi one-dimensional (1D) Bose systems, where the atomic gas is strongly confined in two directions, lead to formation of the Tonks-Girardeu gas whose elementary excitations obey Fermi-Dirac statistics~\cite{Olshanii:98}. In this system, as observed in an optical lattice system of tightly confined 1D tubes~\cite{Paredes:04}, a resonance in the effective 1D coupling strength occurs, leading to strong repulsive interactions so that exchange of atoms is suppressed~\cite{Olshanii:98}. Recently, the origin of this resonance has been attributed to a Feshbach-like, confinement-induced resonance that is associated with a bound state of a closed scattering channel in the transverse mode~\cite{Bergeman:03}. This resonance is of particular importance since it allows for the control of the effective 1D interaction strength and the exploration of different regimes of the BEC vs the Tonks gas by simply varying the confinement of atoms in the transverse direction. 

Concurrently with the exploration of these confinement-induced resonances, we have predicted a ``trap-induced resonance"~\cite{Stock:03} (TIR) for two atoms in separated, tightly confining potential wells, such as the sites of a polarization gradient lin-$\theta$-lin optical lattice~\cite{Deutsch:98}.  This phenomenon arises when a weakly bound molecular state of the dimer is shifted by the potential energy (AC Stark shift) of the separated wells into resonance with a vibrational eigenstate of the trap.  In contrast, optical Feshbach resonances~\cite{Bohn:all}, also induced by off-resonant laser couplings, arise due to strong mixing of excited-state bound molecules with the ground unbound states and are generally accompanied by inelastic processes. Here, the strong gradients in the ac-Stark effect produced by confining optical potentials is sufficient to make ground-state molecules resonant with atomic trap eigenstates, but with tunability different from the magnetic Feshbach resonances arising from the Zeeman shift.  The TIR thus provides an alternative handle for coherently controlling two-atom ground-state interactions.   

In this paper, we evaluate the feasibility of observing these resonances in a particularly promising species, $^{133}$Cs, trapped in optical lattices.  The enormous scattering length of $^{133}$Cs, arising from a very weakly bound molecular state near dissociation, leads to the possibility of creating trap-induced resonances under realistic experimental circumstances.  For a complete description of the atomic interactions, we develop a multichannel scattering model based on the generalized pseudopotential presented in~\cite{Stock:05}. Using this model, we calculate the two-atom energy spectrum under reasonable operating conditions. The remainder of this article is organized as follows.  In Sec. II we review the basic physics of the TIR and its generalization to a multichannel scattering interaction.  We apply this formalism to the specific case of $^{133}$Cs in Secs. III and IV, and conclude in Sec. V.
 

\section{Background}
\subsection{Trap-induced resonances}
\label{TIR}

We review here the salient features of the TIR~\cite{Stock:03}. Consider two interacting atoms in spatially separated harmonic traps, described by the center of mass and relative coordinate Hamiltonians,
\begin{eqnarray}\label{Eq_hamiltonian}
{\hat{H}_{\mathrm{CM}}}&=&\frac{\mat{\hat P}_{R}^{2}}{2M}+\frac{1}{2} M \omega^{2}\mat{R}^{2},\nonumber\\
{\hat{H}_{\mathrm{rel}}}&=&\frac{ \mat{\hat p}_{\mathrm{rel}}^2}{2 \mu} +
\frac{1}{2} \mu \omega^2 \left({\mat{r} - \Delta \mat{z}}\right)^2 +
 \hat{V}_{\mathrm{int}}(\mathbf{r})\, ,
\end{eqnarray}
where $M=2m$ and $\mu=m/2$ are total and reduced masses, respectively.  Though the two atoms may be identical, we assume they are effectively distinguished by their internal states.  The traps for the two atoms are separated in the $z$-direction by $\Delta z$. Since the characteristic length scale of the van der Waals interaction, $\beta_6$, is typically much smaller than the size of the trap ground state~\cite{Blume:02,Bolda:02}, the interatomic interaction $\hat{V}_{\mathrm{int}}(\mathbf{r})$ can be replaced by a contact ``pseudopotential" for $s$-wave scattering,
\begin{equation}
\hat{V}_{\mathrm{eff}}(\mathbf{r}, E_K) = \frac{2 \pi \hbar^2}{\mu}
a_{\mathrm{ eff}}(E_K) \delta(\mathbf r) \frac{\partial}{\partial r}r \,.
\end{equation}
Note, that we parametrize the pseudopotential with an energy-dependent scattering length $a_{\mathrm{ eff}}(E_K)$~\cite{Stock:03,Bolda:02}.  This is particularly important for strong interactions outside the Wigner threshold regime.  In addition, such a pseudopotential not only describes the complete energy-dependent scattering behavior, but also the bound state spectrum if the scattering length is analytically continued to negative energies and the Hamiltonian's eigenvalues are found self-consistently. 

The energy spectrum associated with Eq. (\ref{Eq_hamiltonian}) is easily understood in two limits.  When $\Delta z \rightarrow \infty$, interactions are negligible and the spectrum is a harmonic ladder in three dimensions.  When $\Delta z \rightarrow 0$, we map to the problem of two atoms in a harmonic trap, interacting via a $\delta$-function potential in three dimensions, parametrized by the scattering length $a$.  The spectrum of this problem was solved by Busch {\em et al.}~\cite{Busch:98} and generalized to higher partial wave interactions in~\cite{Kanjilal:04,Peach:04,Stock:05}, revealing positive (negative) energy shifts of the harmonic ladder for $a>(<)0$ that saturate at $\hbar \omega$ for $\left| a \right| \rightarrow \infty$.  In addition, for $a>0$, the spectrum includes a perturbed bound-state of the $\delta$-function potential. This state represents the near-dissociation molecular bound state associated with the positive scattering length of an attractive potential.  

For intermediate separations $\Delta z$, we treat the term $( \mu \omega^2 \Delta z) r \cos \theta$ as a perturbation to the Busch solution and rediagonalize the Hamiltonian. This is done for a range of scattering lengths to determine $E(a)$.  We then solve {\em independently} the problem of the scattering length as a function of kinetic energy, $a(E_K)$ for a given {\em realistic} interaction potential between the particles in free space.  We numerically search for the self-consistent solution, setting $E_K = E(a) - V_{\Delta z}$, where $V_{\Delta z}= \mu \omega^2 \Delta z^2/2$ is the potential energy the two particles must overcome to collide at zero range. This procedure allows us to separate the long-range behavior of the wave function from its short-range behavior and then match them in one self-consistent solution. This is particularly important for solving realistic atom interaction where molecular potentials have the range of angstroms whereas the trap scale is on the order of many nanometers.
\begin{figure}[htbp]
\centerline{\includegraphics[width=85mm]{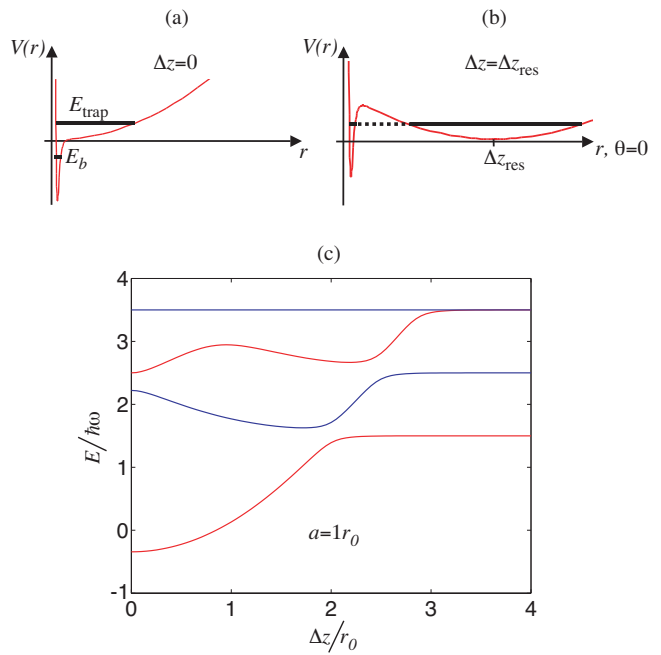}}
\caption[Schematic of trap-induced resonance.]{Schematic of trap-induced resonance and eigenspectrum as a function of separation. Fig. (a) shows the sum of the harmonic trapping potential and chemical binding potential (gray line) in the relative coordinate $z$ for zero trap separation. Fig. (b). The molecular bound state at $E_b$ and trap eigenstate at $E_\mathrm{trap}$ can become resonant at a critical separation $\Delta z_\mathrm{res}$. Fig. (c). Eigenspectrum for two interacting atoms in separated traps for a constant scattering length of $a=1 r_0$ ($r_0=\sqrt{\hbar/\mu\omega}$). Avoided crossings occur in the eigenspectrum as a function of separation $\Delta z$ for all vibrational state with zero transverse excitation.}\label{Fig_TIR}
\end{figure}

To verify the use of the energy-dependent pseudopotential approximation, we have employed a spherically symmetric step potential as a toy model of the molecular binding~\cite{Stock:03}.  For this problem, we can find an analytic solution for the scattering phase-shift from which we deduce the energy-dependent scattering length.  The self-consistent solution is compared with the full numerical solution to the toy problem for a potential with large positive scattering lengths at zero energy. For atoms in very tight traps, as the separation is increased, the trapping potential can be strong enough to raise a bound state of the molecular interaction to positive energies [see Figs.~\ref{Fig_TIR}(a)-\ref{Fig_TIR}(b)]. Avoided crossings occur in the eigenspectrum for certain separations where the energy of this molecular bound state becomes resonant with the eigenstates of the trap potential [see Fig.~\ref{Fig_TIR}(c)]. This effect generally occurs if the interaction potential for the free atoms possesses a molecular bound state very close to dissociation, i.e., the scattering length in free space is positive and on the order of or larger than the size of the trap ground state.  Here, the self-consistent solution, which employs an energy-dependent scattering length, is crucial; a constant scattering length approximation does not capture the location of the bound state and the resulting energy gap accurately. Also note that in this case, analytic continuation of the scattering length to negative energies is essential.  The TIR occurs through a tunneling barrier as the trap distorts the atomic potential at distances large compared to the range of the molecular binding.  As such we probe the ``negative energy" spectrum of the interaction (with respect to collisions of free particles) and the scattering length used for the self-consistent solution is taken at the, usually negative, kinetic energy in the system.

\subsection{The multichannel pseudopotential approximation}
The model presented thus far involves only single channel elastic scattering. A complete analysis of the TIR for a given atomic species requires a generalization of the pseudopotential to multiple scattering channels, including the full spin-dependent nature of the cold-collision process via the hyperfine and exchange interactions, and higher-order partial-wave scattering. We thus require a multichannel matrix generalization of the scattering length that parametrizes the pseudopotential operator, thus allowing the inclusion of both elastic and inelastic processes. We accomplish this by solving the Schr\"{o}dinger equation for two interacting particles in free space in a standard close-coupling expansion of the total wave function with good quantum numbers $(J,M,\Pi)$, the total angular momentum, its projection on the internuclear axis, and parity respectively~\cite{Mies:80},
\begin{equation}
\Psi_\gamma(J,M,\Pi,E)=\sum_\alpha \psi_{\gamma'}(J,M,\Pi;\mat{r}_{el},\hat{r}) \frac{F_{\gamma',\gamma}(r)}{r}\,.
\end{equation}
Here, the indices $\gamma$ and $\gamma'$ refer to the different ``in'' and ``out'' channels of stationary scattering theory~\cite{Taylor}. We choose the channel-state basis to the product basis of atomic hyperfine sublevels for the internal degree of freedom,  $\ket{f_1,m_{f1}}\ket{f_2,m_{f2}}$, plus the partial wave quantum numbers for the relative motion, $\ket{l,m}$. The functions, $F_{\gamma',\gamma}(r)$, are reduced radial wave functions that satisfy the close-coupled equations,
\begin{multline}\label{Eq_closecoupled}
\left( \frac{d^2}{d r^2} -\frac{l'(l'+1)}{r^2} + k_{\gamma'}^2 \right) F_{\gamma',\gamma}(r) \\ = \sum_{\gamma''}\frac{2\mu}{\hbar^2}\left(\hat{V}_{\gamma',\gamma''}(r)-E_{\gamma'}^\infty \delta_{\gamma',\gamma''}\right) F_{\gamma",\gamma}(r) \,,
\end{multline}
with wave number $k_\gamma^2={2\mu} (E-E_\gamma^\infty)/{\hbar^2}$ for the asymptotic channel state energy $E_\gamma^\infty$. In the asymptotic limit where $\hat{V}_{\gamma', \gamma} \rightarrow 0$, the equations decouple and the solutions are linear combinations of the  (reduced) spherical Bessel functions defined as $J_{\gamma}(r) \equiv \sqrt{k_\gamma} r j_l (k_{\gamma} r)$ and Neumann functions $N_{\gamma}(r) \equiv  \sqrt{k_\gamma} r n_l (k_{\gamma} r)$ ,
\begin{equation}\label{Eq_closeasym}
F_{\gamma',\gamma}^{\infty}(r) \rightarrow J_{\gamma'}(r)\delta_{\gamma',\gamma}\ + N_{\gamma'}(r) K_{\gamma',\gamma}\, ,
\end{equation}
expressed in terms of the $K$ matrix~\cite{Taylor}.  Typically, the scattering length is defined by the scattering phase-shift via the $S$ matrix. As we will see, however, in the pseudopotential approximation described below, it is advantageous and more appropriate to define a scattering length matrix through the $K$ matrix. As derived in detail below, basing the definition of the pseudopotential on the $K$ matrix allows us to correctly match the scattering boundary conditions for the asymptotic wave functions in Eq.~(\ref{Eq_closeasym}). In addition, note that our definition of the reduced Bessel functions differs in phase convention from~\cite{Mies:80}; for closed channels these functions are purely imaginary.

The objective of the pseudopotential formalism is to replace the exact atomic interaction potential at all energies, for all partial waves, and for all channels by a zero-range pseudopotential. In~\cite{Stock:05}, we showed how such a pseudopotential could be constructed for an arbitrary partial wave using the limit of a $\delta$ shell of radius $s$, in the limit $s \rightarrow 0$. We generalize here to a multichannel version of this pseudopotential, including anisotropic couplings. We define an energy-dependent, $l$-wave scattering length matrix $a_{\gamma,\gamma'}^{l+l'+1}(k_{\gamma},k_{\gamma'})$ in terms of the multichannel $K$ matrix,
\begin{equation}\label{Eq_a_matrix}
k_{\gamma}^{l+1/2} k_{\gamma'}^{l'+1/2}a_{\gamma',\gamma}^{l+l'+1}(k_{\gamma},k_{\gamma'})=K_{\gamma',\gamma}\,.
\end{equation}
A pseudopotential with an {\it energy-dependent} scattering length matrix has the same properties as the single-channel pseudopotential, capturing the complete scattering and bound-state properties of the participating channels~\cite{Stock:03,Stock:05}. It will further allow us to continue the scattering-length matrix element for the channel of interest to negative energies, which are crucial for the case of atoms in separated traps (see Sec. III). 

Using the definition in Eq.~\ref{Eq_a_matrix}, we take as our ansatz for the multichannel pseudopotential
\begin{equation}\label{Eq_multi_ansatz}
\hat{V}(r) = \sum_{\gamma,\gamma'}  \ket{l,m} \hat{V}_{\gamma, \gamma'}(r) \bra{l',m'}
\end{equation}
with 
\begin{equation}\label{Eq_multi_pseudo}
\hat{V}_{\gamma, \gamma'}(r)=  c_{l,l'}a_{\gamma',\gamma}^{l+l'+1} \frac{\delta(r-s)}{r^{l+l'+2}}  \hat{P}_{l'}\, ,
\end{equation}
where the constants $c_{l,l'}$ are to be determined.  The regularization operator,
\begin{equation}
\hat{P}_l = \frac{r^{l+1}}{(2l+1)!} \frac{\partial^{2l+1}}{\partial r^{2l+1}} r^l \,,
\end{equation}
acts near the origin as the identity on regular radial wave functions $J_{\gamma}(r)$ and the null operator on the irregular function $N_{\gamma}(r)$.  Please note that in contrast to~\cite{Stock:05}, we define here the pseudopotential to act on the {\em reduced} radial wave functions. The solution to the radial equation for the pseudopotential has the form
\begin{eqnarray}\label{Eq_radialsolution}
F_{\gamma',\gamma}^{>} =& J_{\gamma'}(r)\delta_{\gamma',\gamma} + N_{\gamma'}(r)K_{\gamma',\gamma}\,, & r>s.\nonumber\\
F_{\gamma',\gamma}^{<} =&  J_{\gamma'}(r) A_{\gamma',\gamma}\,,  & r<s.
\end{eqnarray}
The matrix $A_{\gamma',\gamma}$ follows from continuity of the wave function across the shell, 
\begin{equation}\label{coefficient}
A_{\gamma',\gamma}=\delta_{\gamma',\gamma} -\frac{(2l'-1)!! (2l'+1)!!}{(k_{\gamma'}s)^{2l'+1}} K_{\gamma',\gamma}\,.
\end{equation}
Inserting these solutions and the ansatz in Eq.~(\ref{Eq_multi_ansatz}) into the close-coupled equation~(\ref{Eq_closecoupled}) and leaving the limit $s\rightarrow0$ for later, we need to evaluate the action of the pseudopotential on the wave function,
\begin{eqnarray}\label{pseudo_act}
&&\hat{V}_{\gamma',\gamma''}(r) F_{\gamma'',\gamma}^{>}(r) \nonumber \\
&&= \left[c_{l',l''} a_{\gamma'',\gamma'}^{l'+l''+1}\frac{\delta(r-s)}{r^{l'+l''+2}}  \right] \hat{P}_{l''}  F_{\gamma'',\gamma}^{>}(r) \nonumber\\
&&=\left[c_{l',l''}  a_{\gamma'',\gamma'}^{l'+l''+1}\frac{\delta(r-s)}{r^{l'+l''+2}}  \right] \left[\frac{k^{l''+1/2}_{\gamma''} r^{l''+1}}{(2l''+1)!!}  \delta_{\gamma'',\gamma} \right],
\end{eqnarray}
having applied the regularization operator near the origin.  The remaining boundary condition is found by integrating the close-coupled equations around the shell,
\begin{multline}\label{multishell_integral}
\lim_{\epsilon\rightarrow0}\left( \left[ \frac{d F_{\gamma',\gamma}^{>}(r)}{d r} \right]_{r=s+\epsilon} -   \left[ \frac{d F_{\gamma',\gamma}^{<}(r)}{d r} \right]_{r=s-\epsilon} \right)\\
= \lim_{\epsilon\rightarrow0}\int^{s+\epsilon}_{s-\epsilon}  \sum_{\gamma''} \hat{V}_{\gamma',\gamma''}(r) F_{\gamma'',\gamma}^{>}(r)  \,.
\end{multline}
Using Eqs. (\ref{Eq_radialsolution}), (\ref{coefficient}), and (\ref{pseudo_act}), we find
\begin{multline}
K_{\gamma',\gamma} \frac{(2l'+1)!!}{k_{\gamma'}^{l'+1/2} s^{l'+1}}  \\=  \sum_{\gamma''} \left[c_{l',l''}  \frac{a_{\gamma'',\gamma'}^{l'+l''+1}}{s^{l'+l''+2}}\frac{k^{l''+1/2}_{\gamma''} s^{l''+1}}{(2l''+1)!!} \delta_{\gamma'',\gamma} \right] \,.
\end{multline}
Since the $K$ matrix is symmetric $K_{\gamma',\gamma}=K_{\gamma,\gamma'}$~\cite{Mies:80}, the close-coupled equation set is fulfilled and using Eq. (\ref{Eq_a_matrix}) we find the strength of the pseudopotential is $c_{l',l''}=(2l'+1)!!(2l''+1)!!$. In summary, the pseudopotential (defined as acting on the reduced wave functions) is
\begin{equation}\label{pseudopotential}
\hat{V}_{\gamma, \gamma'}(r)=  \frac{(2l+1)!!(2l'+1)!!}{(2l'+1)!} \,a_{\gamma',\gamma}^{l+l'+1}\, \frac{\delta(r-s)}{s^{l+1}} \frac{\partial^{2l'+1}}{\partial r^{2l'+1}}r^{l'}\,.
\end{equation}
We apply this solution to the case $^{133}$Cs in the next section.

\section{Calculation of the scattering length matrix of Cesium for positive and negative energies}
\label{Cs_scattering_length}

The cold collision properties of $^{133}$Cs are anomalous due to the existence of a bound state of the $^3 \Sigma^+_u$ molecular potential very close to dissociation~\cite{Leo:00} and the large second-order spin-orbit coupling~\cite{Chin:04}. Whereas the former property leads to extremely large triplet scattering lengths and possible trap-induced resonances, the latter implies possible strong inelastic collision channels.  As such, a realistic modeling of the scattering is necessary to study the feasibility of observing the TIR. We will accomplish this using the numerical solution to the close-coupling equation~(\ref{Eq_closecoupled}) developed at the National Institute for Standards and Technology~\cite{NIST:codes} based on the best known cesium dimer interaction potentials, recently refined through spectroscopic measurements of Feshbach resonances~\cite{Chin:04}.  The multichannel scattering solutions will allow us to determine the energy-dependent scattering-length matrix as input to the pseudopotential in Eq.~(\ref{Eq_multi_pseudo}).

\begin{figure}[htbp]
\centerline{\includegraphics[width=80mm]{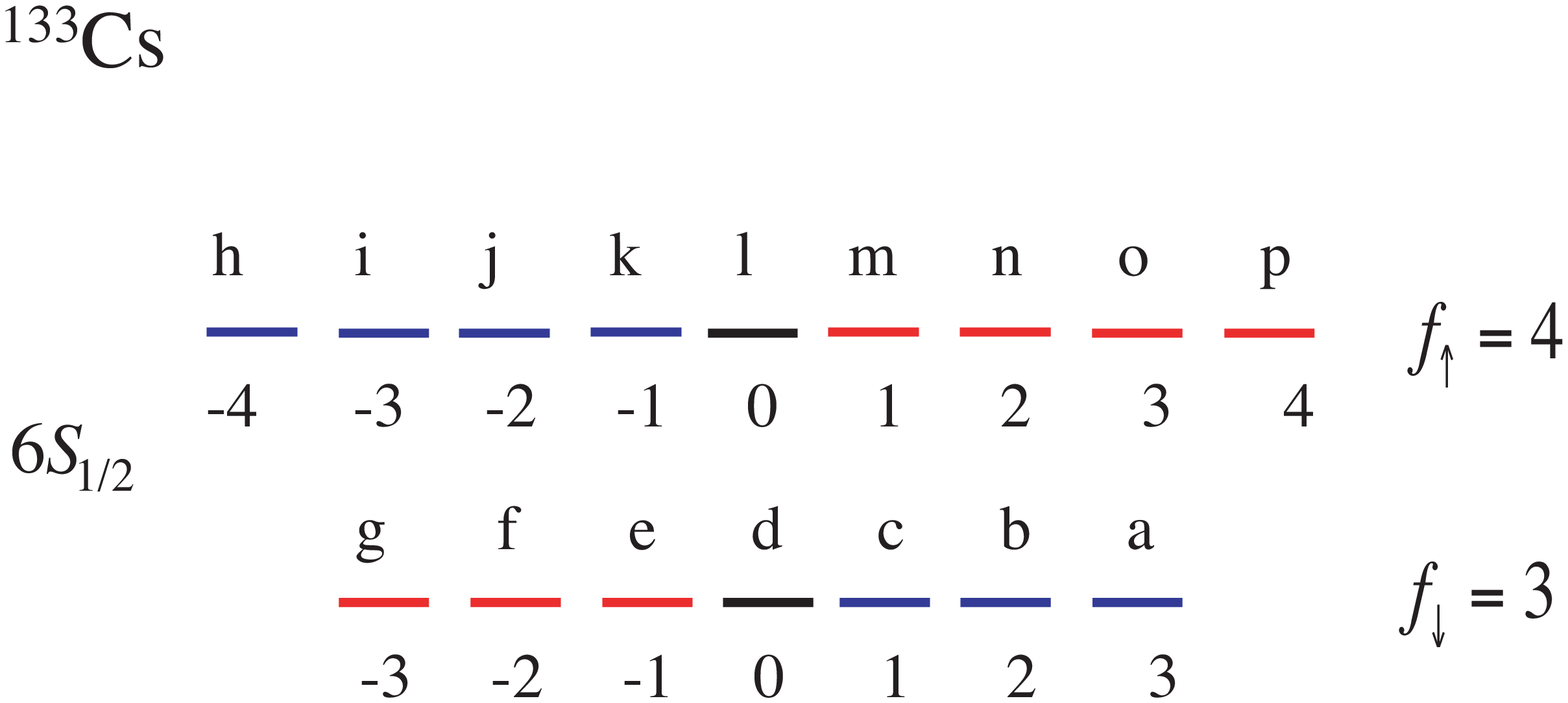}}
\caption[Hyperfine levels of the $6S_{1/2}$ ground state of $^{133}$Cs.]{Hyperfine levels of the $6S_{1/2}$ ground state of $^{133}$Cs. The magnetic sublevels are labeled with letters from a to p.}\label{Fig_Cs_hyperfine}
\end{figure}
Figure~\ref{Fig_Cs_hyperfine} shows the atomic hyperfine states and magnetic sublevels in the $^{133}$Cs atomic ground state labeled with letters from $a$ to $p$. The scattering channel of interest is the two-atom combination in the stretched states, $\ket{ap}$, with a total angular momentum projection quantum number of $m_\mathrm{total}=m_{f1}+m_{f2}=7$. This channel is of particular interest since, in the absence of any spin-motion coupling, conservation of angular momentum implies conservation of these quantum numbers. For this reason, the equivalent of states $a$ and $p$ for $^{87}$Rb were the ones originally considered for entangling atoms via ultracold ground-state collisions~\cite{Jaksch:99}. For $^{133}$Cs atoms, this desirable property is not automatically satisfied due to the higher-order spin-orbit interaction~\cite{Leo:00,Chin:04}, which couples a multitude of open and closed scattering channels.  In our calculation, 49 channels were included in the close-coupling set (see Table~\ref{Table_channels}).
\begin{table}
\centerline{\includegraphics[width=85mm]{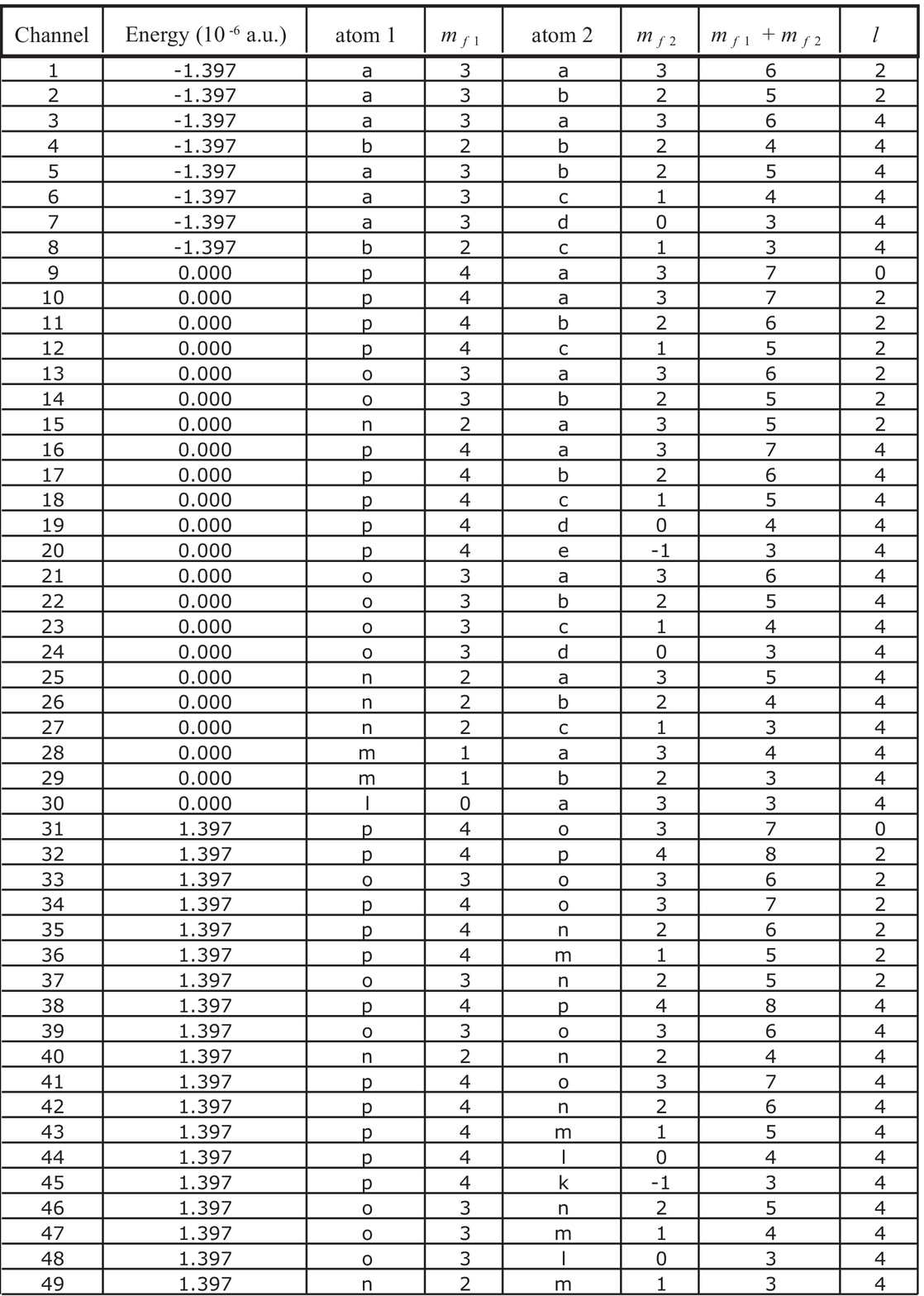}}
\caption[Participating channel information.]{Participating channel information. The channel of interest is $\ket{ap}$ (channel 9), which defines the zero of energy (in atomic units a.u.). The first 30 channels are open, of which the first eight channels are the ones with both atoms in the lower hyperfine manifold. Channels 31 to 49 are closed channels with both atoms in the $f=4$ manifold. The channel states are labeled following the conventions of Fig.~\ref{Fig_Cs_hyperfine} (columns 3 and 5). The table includes the magnetic quantum numbers $m_f$ for each atom (columns 4 and 6), the total magnetic quantum number $m_\mathrm{total}$ (column 7) and the partial wave quantum number $l$ (last column).}\label{Table_channels}
\end{table}

The set of close-coupled equations for the reduced radial wave functions of the participating channels is integrated from an initial radius $r_\mathrm{i}$ to a final radius $r_f$ using a standard renormalized Numerov method~\cite{Johnson:77,Johnson:78} for both open and closed channels. There, the logarithmic derivative matrix $m_{\gamma',\gamma}(r_f)$ is calculated for the reduced wave functions $F_{\gamma',\gamma}(r)$ according to
\begin{equation}
m_{\gamma',\gamma}(r_f)=\frac{ \frac{\partial}{\partial r} F_{\gamma',\gamma}(r)\vert_{r=r_f}}{F_{\gamma',\gamma}(r_f)}\,.
\end{equation}
Matching the logarithmic derivative to the asymptotic wave function, we can determine the $K$-scattering matrix from $m_{\gamma',\gamma}(r_f)$ according to
\begin{multline}\label{Eq_logdermat}
K_{\gamma',\gamma}  = \left[ N_{\gamma'}(r_f) m_{\gamma',\gamma}(r_f) - N'_{\gamma'}(r_f)\right]^{-1}\\ \left[ J_{\gamma'}(r_f) m_{\gamma',\gamma}(r_f) - J'_{\gamma'}(r_f)\right]\,.
\end{multline}
Here, $J'_{\gamma'}(r_f)$ and $N'_{\gamma'}(r_f)$ are the derivatives of the reduced spherical Bessel functions at $r=r_f$. The $K$ matrix then defines the scattering length matrix that is appropriate for use in the generalized multichannel pseudopotential as discussed in the previous section. Note that in order to get a convergent value of the $K$ matrix, the final integration radius, $r_f$, can be larger than the typical size of the ground state in an optical lattice site. This might raise doubts about the validity of the pseudopotential approximation in an optical lattice system, since the effective scale of the interaction seems to be much larger than the trap size. However, as shown in Refs.~\cite{Bolda:02,Blume:02}, it is the characteristic $\beta_6$ length scale that affects the validity of the pseudopotential approximation.  As long as the trapping potential does not distort the chemical binding potential, it is valid to calculate the energy-dependent scattering length in {\em free space} and use this as the input to the self-consistent solution in the trap. Since $\beta_6=100a_0$ in $^{133}$Cs, and therefore much smaller than typical trap sizes in tight optical lattices, we can safely approximate the interaction by a  pseudopotential. 

\begin{figure}[t]
\centerline{\includegraphics[width=80mm]{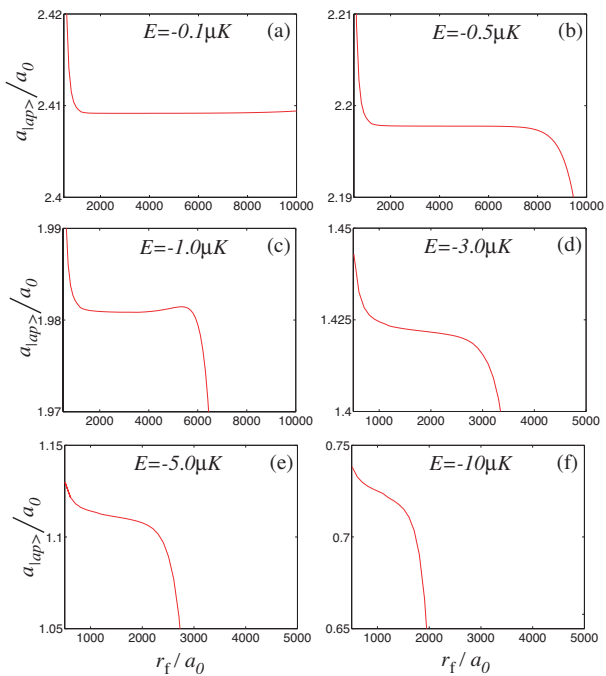}}
\caption[Calculations of the diagonal scattering length matrix element for the $\ket{ap}$ channel as a function of the stopping point $r_f$.]{Calculations of the diagonal scattering length matrix element for the $\ket{ap}$ channel as a function of the stopping point $r_f$ for different negative energies.}\label{Fig_convergence_test}
\end{figure}

\begin{figure}[htbp]
\centerline{\includegraphics[width=80mm]{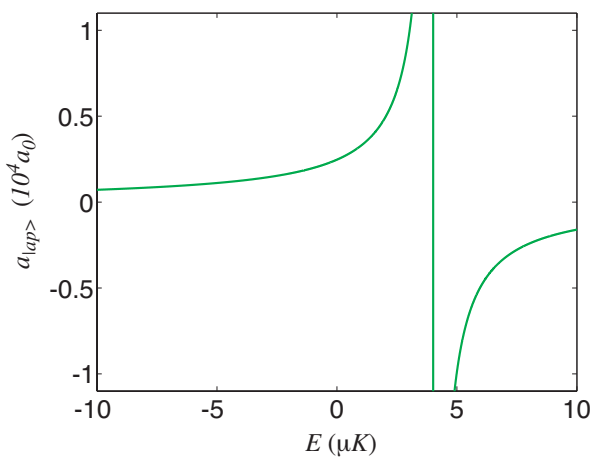}}
\caption[Calculations of the diagonal scattering length matrix element for the $\ket{ap}$ channel as a function of energy.]{Calculations of the diagonal scattering length matrix element, derived from the $K$ matrix, for the $\ket{ap}$ channel as a function of energy.}\label{Fig_ap_scatteringlength}
\end{figure}

As discussed in Sec.~\ref{TIR}, negative energy scattering becomes important for two atoms in nonoverlapping traps due to the tunneling barrier created by the trapping potential. In this case, the scattering channels of interest are closed and we must extend the $K$ matrix to include these channels. This can be achieved by simply continuing the reduced spherical Bessel functions $J_{\gamma}(r_f)$ and $N_{\gamma}(r_f)$ to negative energies by using a purely imaginary wave vector, $k=i\kappa$, and allowing a complex argument of the Bessel functions (note, in our convention for the reduced Bessel function, these functions are then pure imaginary). The complete $K$ matrix, which is Hermitian for open channels and anti-Hermitian for closed channels, is calculated using Eq.~(\ref{Eq_logdermat}).  Figures~\ref{Fig_convergence_test}(a)-\ref{Fig_convergence_test}(f) show the diagonal scattering length matrix element for the $\ket{ap}$ channel as a function of the stopping point $r_f$ for various negative energies. At negative energies, we expect the codes to converge in the calculation of $a_{\ket{ap}}$ only for a small range of  $r_f$. The close-coupled equations need to be integrated far enough into the asymptotic regime where the interaction potential vanishes. However, for very large $r$, we expect the wave function to diverge exponentially and the numerical Numerov codes to become unstable. We expect the region of convergence to decrease for larger negative energies, since the wave function diverges as $\exp{(+\kappa r)}$ (see Figs.~\ref{Fig_convergence_test}).  
As seen in Figs.~\ref{Fig_convergence_test}, we can reliably calculate the $\ket{ap}$ scattering length matrix element for negative energies as low as $-10\,\mu K$. The calculation errors in the scattering length of Fig.~\ref{Fig_ap_scatteringlength} can be estimated from Figs.~\ref{Fig_convergence_test} to be around $5\%$ at $-10\,\mu K$ whereas for smaller negative energies of around $-1\,\mu K$ the error can be estimated to be smaller than $0.5\%$.

\begin{figure}[htbp]
\centerline{\includegraphics[width=85mm]{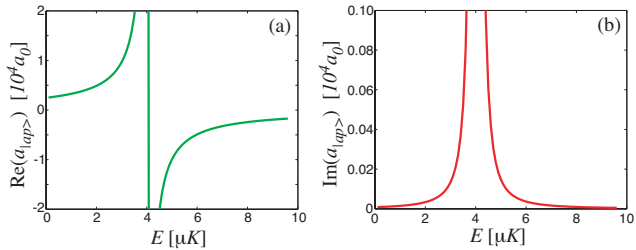}}
\caption[$S$ matrix-based calculations of the real and imaginary part of the scattering length.]{$S$ matrix based calculations of the real (a) and imaginary part  (b) of the scattering length for the $\ket{ap}$ channel as a function of energy.}\label{Fig_a_imag}
\end{figure}

A plot of the diagonal scattering-length element as a function of energy for the $\ket{ap}$ channel is shown in Fig.~\ref{Fig_ap_scatteringlength}. The curve continues smoothly across zero energy and also takes a finite value at the bound-state energy as expected. 
At a positive energy, $E\approx 4\,\mu K$, there is a resonance in the scattering length that is associated with inelastic processes. Examining the scattering-matrix off-diagonal elements, the $\ket{ap}\ket{l=0}$ channel primarily couples to higher partial wave channels in the same hyperfine manifold, channels 10-30 in Table~\ref{Table_channels}, and in particular to the $\ket{ap}\ket{l=2}$ and $\ket{ao}\ket{l=2}$ channels. Exothermic couplings that result in both atoms in the lower hyperfine manifold, e.g., couplings to the $\ket{aa}\ket{l=2}$ channel, are one to two orders of magnitude smaller. To better characterize the off-diagonal couplings to {\em{all}} channels, we will adopt a complex scattering length approach in the following. While the $K$ matrix describes the asymptotic boundary conditions and is therefore essential in the pseudopotential description, the scattering $S$ matrix is a measure of the coupling between different channels. These inelastic processes and the resulting loss from the channel of interest are therefore more appropriately described by a complex scattering length defined through the $S$ matrix, which is related to the $K$ matrix by,
\begin{equation}\label{Eq_smatrix}
\mat{S}\equiv\left[\mat{1} +i \mat{K} \right]\left[\mat{1} - i  \mat{K} \right]^{-1}\,.
\end{equation}
By unitarity the diagonal $S$ matrix elements are
\begin{equation}
S_{\gamma,\gamma}=1-\sum_{\gamma'}S_{\gamma',\gamma}=\exp{\left[ 2 i \left(\lambda + i \mu \right) \right]}=\abs{S_{\gamma,\gamma}} e^{ 2 i \lambda}\,,
\end{equation}
where we have defined a complex phase shift $\delta=\lambda + i \mu$. This can be used to define a complex energy-dependent scattering length~\cite{Bohn:97},
\begin{equation}
\tilde{a}_{\gamma}=a_{\gamma}-i\beta_{\gamma}\equiv-\frac{  \tan{\left(\lambda+i\mu\right)}}{k_{\gamma}^{2l+1}}\,.
\end{equation}
The imaginary part of the scattering length is a measure of the ``loss'' from the scattering channel $\gamma$ to {\em{all}} other open channels. Note that the imaginary part of the scattering length not only measures the inelastic collisions but also includes the coherent couplings to channels in the same hyperfine manifold. Figure~\ref{Fig_a_imag} shows the real and imaginary part of the complex scattering length for positive energies. For scattering energies below $4\,\mu K$, the imaginary part is several orders smaller than the real part of the scattering length. In this regime, couplings to other channels that are due to spin-spin or spin-orbit coupling can be neglected and the total magnetic quantum number $m_\mathrm{total}$ is approximately conserved.  This implies that, in this energy regime, interactions of atoms in $\ket{ap}$ can be treated to good approximation as single channel scattering. 

We have also examined the magnetic field dependence of the scattering length in the linear Zeeman regime.
\begin{figure}[htbp]
\centerline{\includegraphics[width=80mm]{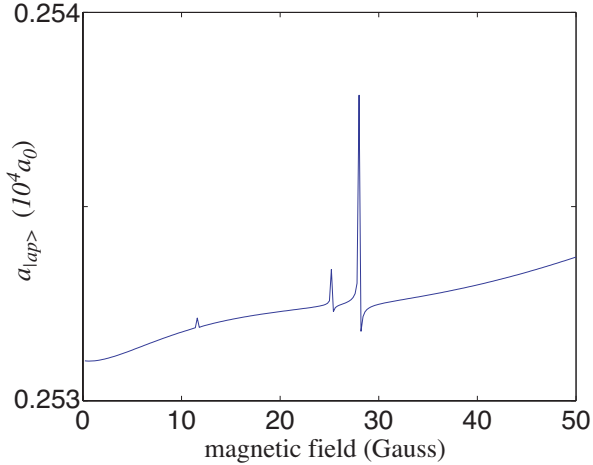}}
\caption[Calculations of the diagonal scattering-length matrix element for the $\ket{ap}$ channel as a function of the magnetic field.]{Calculations of the diagonal scattering length matrix element for the $\ket{ap}$ channel as a function of magnetic field. Three narrow scattering resonances can be observed for applied magnetic fields in the range from $0$ to $50$ G.}\label{Fig_B_field_dependence}
\end{figure}
The background scattering length remains large and constant over a wide range of magnetic fields and shows three narrow resonances over the calculated range at $11.6$, $25$, and $28$ G in Fig.~\ref{Fig_B_field_dependence}. An identification of the different resonances, i.e., the labeling of the resonances by the corresponding bound-states, is left for future investigation. The resonances are very narrow (less than $1$ G) and should not lead to any strong magnetic field sensitivity in the atomic interactions.

\section{Trap-induced shape resonances of Cesium in optical lattices}

We consider a three dimensional (3D) optical lattice, created by three pairs of linearly polarized counter-propagating laser beams. Along one direction we allow for polarization gradients in a lin-angle-lin configuration~\cite{Deutsch:98} which shifts $\sigma+$ and $\sigma-$ standing waves and carries atoms in appropriate sublevels together from neighboring sites and probes the TIR. We approximate the optical lattice wells by harmonic oscillator potentials with frequency $\omega$ with 
\begin{equation}
\hbar\omega=2\sqrt{V_{pp} E_R}\,.
\end{equation}
$V_{pp}$ is the peak-to-peak depth of the optical lattice potential and $E_R$ is the recoil energy of a $^{133}$Cs atom.  The Lamb-Dicke parameter, which measures the localization of atoms in the lattices, $\eta=\sqrt{E_R/\hbar\omega}=k_l \bar{r}_0$, where $\bar{r}_0=\sqrt{\hbar/2m\omega}$ is the full width at half maximum (FWHM) of the harmonic-oscillator ground state of a single atom (mass m). As we deal with relative motion of two atoms, we choose to scale in terms of characteristic harmonic oscillator units  $r_0=\sqrt{\hbar/\mu\omega}= r_0=2 \bar{r}_0$. 

For the observation of a TIR for the $\ket{ap}$-channel, we require $r_0$ to be on the order of, or much smaller than, the scattering length which takes the value $a_{\ket{ap}}\approx2500a_0\approx 132$~nm in the negative energy regime of interest.  In addition, in order to induce a TIR, the trap potential at the origin, $V_{\Delta z}=\frac{1}{2}\Delta z^2$, must be strong enough to raise the molecular bound-state at $E_b$ to positive energies so that $E_b+V_{\Delta z} \geq \frac{3}{2}\hbar\omega$. In an optical lattice $V_{\Delta z}$ is limited by $V_{pp}$, resulting in the constraint $\frac{3}{2}\hbar\omega<V_{\Delta z}<V_{pp}$. An optical lattice with a Lamb-Dicke parameter of $\eta=0.25$, $V_{pp}=64E_R$,  $\hbar\omega=16 E_R$, and $\bar{r}_0\approx34$~nm fulfills all these requirements and still allows a reliable harmonic approximation to the optical lattice potential. Also, for $^{133}$Cs the $\beta_6$ parameter is on the order of $100 a_0$ or $5.3$ nm, i.e., much smaller than $\bar{r}_0$, therefore allowing a reliable approximation of the interaction via a $\delta$-shell pseudopotential.

A complete description of interacting atoms in separated traps should include a calculation of the energy spectra as a function of trap separation for the scattering channel of interest, as well as for all channels with off-diagonal couplings to this channel. It furthermore should include the state-dependent nature of the optical lattice potentials. While this is of great importance for most scattering channels, for the $\ket{ap}$ scattering channel, as shown in the previous section, the off-diagonal couplings can be safely neglected below $4\,\mu K$. The calculation of the energy spectrum for the $\ket{ap}$ channel using only a {\em single channel} in Eq.~(\ref{pseudopotential}) with $l,l'=0$ is sufficient as long as the appropriate scattering length matrix element calculated from the $K$ matrix is used to characterize the interaction. 

\begin{figure}[htbp]
\centerline{\includegraphics[width=80mm]{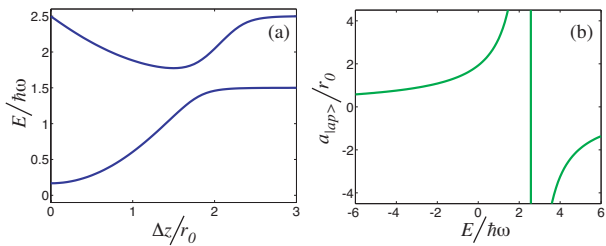}}
\caption[Self-consistent energy eigenvalues as a function of trap separation for collisions of $^{133}$Cs atoms.]{(a) Self-consistent energy eigenvalues as a function of trap separation $\Delta z$ calculated using the $K$ matrix based scattering length for the $\ket{ap}$-channel in $^{133}$Cs, which is shown in (b) in harmonic oscillator units. }\label{Fig_Cs_energy}
\end{figure}

Figure~\ref{Fig_Cs_energy}(a) shows the calculated energy spectrum for $^{133}$Cs atoms in separated harmonic isotropic traps as a function of trap separation $\Delta z$, assuming an optical lattice with a Lamb-Dicke parameter of $\eta=0.25$. The energy eigenvalues are calculated self-consistently at each separation as described in Sect.~\ref{TIR} using the energy-dependent scattering length for $^{133}$Cs, which is shown in Fig.~\ref{Fig_Cs_energy}(b) in harmonic oscillator units. This plot of the scattering length nicely illustrates the necessity of the application of an {\em energy-dependent} pseudopotential because of the strong energy dependence of the scattering length in the energy range of interest. A constant scattering-length approximation would misestimate the location of the bound state and therefore the size of the avoided crossing in the energy spectrum.
Figure~\ref{Fig_Cs_energy}(a) shows the main signature of the trap-induced resonance in the form of the avoided crossing, which in the case of $^{133}$Cs results in an almost maximal [compare to Fig. 3 (inset) in Ref.~\cite{Stock:03}] energy gap of about $\hbar\omega/2$. 

\begin{figure}[htbp]
\centerline{\includegraphics[width=80mm]{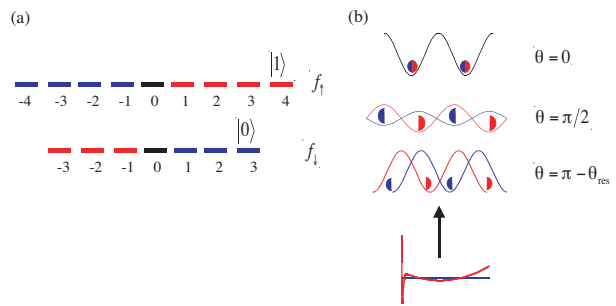}}
\caption[Schematic of controlled collisions via trap-induced resonances]{Encoding in the $^{133}$Cs hyperfine structure and schematic of controlled collisions via trap-induced resonances. (a) Hyperfine levels of the $6S_{1/2}$ ground state of $^{133}$Cs. The logical basis states $\ket{0}$ and $\ket{1}$ are encoded in the stretched states in order to avoid inelastic collisions. (b) Following the proposal by Jaksch~{\it{et al.}}~\cite{Jaksch:99} atoms travel on different lattice potentials. Unlike in the original proposal, here, the phase shift is acquired for separated atoms using the trap-induced resonance.}\label{Fig_TIR_Cs_proposal}
\end{figure}

Controlled ultracold collisions, using two-atom Ramsey interferometry in polarization gradient optical lattices, have been demonstrated by Mandel  {\it{et al.}}~\cite{Mandel:03}. Such experiments can be used to probe the TIR as shown in Fig.~\ref{Fig_TIR_Cs_proposal}. In a linearly polarized lattice, atoms in their vibrational ground state are placed in an equal superposition of states $\ket{a}$ and $\ket{p}$.  These two states are transported in opposite directions as the polarizations are rotated by an angle $\theta$ into a lin-angle-lin configuration. As $\theta$ approaches $180^\circ$ (lin-$\parallel$-lin configuration), neighboring atoms in the $\ket{ap}$ channel collide. The collisional interaction results in a relative energy shift for the two atom state $\ket{ap}=\ket{01}$ which acquires a phase $\phi$ relative to $\ket{00}$, $\ket{10}$, and $\ket{11}$. The interaction entangles these atoms, maximally so when $\phi=\pi$.  In the experiment of Mandel ~{\it{et al.}}, a filled lattice was prepared in a Mott insulator with one atom per site, loaded from a Bose-Einstein condensate~\cite{Greiner:02}.  Controlled collisions lead to an ``entangled chain" of atoms (the 1D cluster state~\cite{Briegel:01all} in a perfectly filled lattice and for $\pi$ phase shift).  The periodic disappearance and reappearance of single atom Ramsey fringes due to this entanglement is a measurement of the scattering phase shift.  For the case at hand, assuming adiabatic transport, this is the phase shift accumulated along the lower energy curve in Fig.~\ref{Fig_Cs_energy}(a).  At a critical separation between the atoms, a large (negative) phase shift will be accumulated, providing a signature of the TIR. By controlling the final separation and hold time of the atoms, one can in principle map the full adiabatic energy curve.

\section{Summary}
In this paper we have considered a trap-induced resonance between a molecular bound state and a trap vibrational state for two $^{133}$Cs atoms in overlapping but separated traps.  To study the interaction between realistic alkali atoms with hyperfine structure, we generalized the $\delta$-shell pseudopotential, introduced in~\cite{Stock:05}, to the case of multichannel scattering parameterized by the $K$ matrix. Using a numerical solution to the close-coupling equations and a self-consistent solution to the Schr\"{o}dinger equation, we calculated the energy spectrum for $^{133}$Cs atoms interacting on the stretched-state channel, $\ket{f=3,m_f=3}\ket{f=4,m_f=4}$, as a function of trap separation. The large energy gap in the spectrum provides a signature by which the trap-induced resonance could be experimentally observed for a lattice with Lamb-Dicke parameter $\eta=0.25$ using Ramsey interferometry to probe the collisional phase shift.

The TIR provides a new method for tuning interactions between ultracold atoms, with applications to quantum degenerate gases, molecular control, and quantum logic. For example, by adiabatically tuning through the TIR to zero separation between neighboring traps, atoms in the first vibrational state are mapped onto the highest molecular bound state. In principle, one can induce coherent superpositions of atomic and molecular states to control new states of matter. The existence of trap-molecule couplings opens avenues for new protocols in quantum control.

An important practical consideration, not treated here, is the effect of finite temperature. As the TIR is explicitly tied to the vibrational state in the trap, the scattering phase shift can be sensitive to motional heating. As seen in Fig.~\ref{Fig_TIR}(c), avoided crossings with the molecular bound state occur for all vibrational trap states associated with zero transverse excitation. For large scattering lengths, considered here, the size of the coupling is not very sensitive to the localization of atoms in the trap and thus a common phase shift should be seen for atoms with either zero or one quantum of longitudinal vibration. This should make the TIR fairly robust as long as the transverse excitation is suppressed. An additional consideration is the effect of spin-dependent trapping.  In our multichannel scattering calculation, we ignored the spin dependent light shift forces that occur in polarization gradient optical lattices. Such forces can close certain channels and provide new mechanisms for robust encodings of quantum logic, immune to spin-flips that would otherwise occur in free space. These considerations will be studied in future research.


\begin{acknowledgments}
We thank Paul Julienne, Carl Williams, Eite Tiesinga, and Sanjiv Shresta for providing the close-coupling codes and very helpful discussions as well as Paul Alsing for his help with the FORTRAN codes. This work was partly supported by NSA/ARDA Contract No.~DAAD19-01-1-0648 and by the ONR Contract No.~N00014-03-1-0508. R. S. also acknowledges support by the AlbertaÕs Informatics Circle of Research Excellence (iCORE).
\end{acknowledgments}


\end{document}